\documentclass[twocolumn,aps,superscriptaddress]{revtex4}%

\usepackage{amsfonts}
\usepackage{amsmath}
\usepackage{amssymb}
\usepackage{graphicx}%
\providecommand{\U}[1]{\protect\rule{.1in}{.1in}}
\providecommand{\U}[1]{\protect\rule{.1in}{.1in}}

\begin{document}
\title{Anisotropic Fabry-P\'{e}rot resonant states confined within nano-steps on the
topological insulator surface}
\author{Zhen-Guo Fu}
\affiliation{Beijing Computational Science Research Center, Beijing 100084, China}
\author{Ping Zhang}
\thanks{zhang\_ping@iapcm.ac.cn}
\affiliation{Institute of Applied Physics and Computational Mathematics, Beijing 100088, China}
\affiliation{Beijing Computational Science Research Center, Beijing 100084, China}
\author{Mu Chen}
\affiliation{Beijing Institute of Aeronautical Materials, Beijing 100095, China}
\author{Zhigang Wang}
\affiliation{Institute of Applied Physics and Computational Mathematics, Beijing 100088, China}
\author{Fa-Wei Zheng}
\affiliation{Institute of Applied Physics and Computational Mathematics, Beijing 100088, China}
\author{Hai-Qing Lin}
\affiliation{Beijing Computational Science Research Center, Beijing 100084, China}

\begin{abstract}
\textbf{The peculiar nature of topological surface states, such as absence of
backscattering, weak anti-localization, and quantum anomalous Hall effect, has
been demonstrated mainly in bulk and film of topological insulator (TI), using
surface sensitive probes and bulk transport probes. However, it is equally
important and experimentally challenging to confine massless Dirac fermions
with nano-steps on TI surfaces. This potential structure has similar ground
with linearly-dispersed photons in Fabry-P\'{e}rot resonators, while reserving
fundamental differences from well-studied Fabry-P\'{e}rot resonators and
quantum corrals on noble metal surfaces. In this paper, we study the massless
Dirac fermions confined within steps along the $x$ ($\Gamma\mathtt{-}$K) or
$y$ ($\Gamma\mathtt{-}$M) direction on the TI surface, and the
Fabry-P\'{e}rot-like resonances in the electronic local density of states
(LDOS) between the steps are found. Due to the remarkable warping effect in
the topological surface states, the LDOS confined in the step-well running
along $\Gamma$-M direction exhibit anisotropic resonance patterns as compared
to those in the step-well along $\Gamma$-K direction, which can be detected by
scanning tunneling microscopy. The transmittance properties and spin
orientation of Dirac fermion in both cases are also anisotropic in the
presence of warping effect.}

\end{abstract}
\maketitle

Recently, the discovery of both two-dimensional and three-dimensional
topological insulators (TIs) \cite{Hasan2010,Qi2011} has attracted enormous
attentions. Unlike the conventional two-dimensional electron states, the
spin-helical surface states of three-dimensional TIs, which are protected by
the time-reversal symmetry and consist of an odd number of spin-helical Dirac
cones, are characterized by the gapless Dirac Hamiltonian. Owing to the chiral
nature of the quasiparticles in TI surface, many unusual effects have been
observed. Among them, the quantum anomalous Hall effect has been theoretically
\cite{FangZ} predicted and experimentally \cite{ChangC} observed in a thin
films of chromium-doped TI material (Bi,Sb)$_{2}$Te$_{3}$, which may open a
door of the application of TI materials in the field of the
low-power-consumption electronics. A robust response of weak antilocalization
\cite{Chen2010,Checkelsky,Liu2012}\ has been tested in the transport studies,
and the suppression of the backscattering has also been confirmed by examining
the scattering from impurities or step edges
\cite{Roushan,ZhangT,Seo,WangJ,Biswas,FuZG,WangZ,ZhangDg,AnJ,Rakyta,Alpichshev}
in previous scanning tunneling microscopy (STM) reports. Particularly, because
of the warping effect \cite{FuL} in topological surface states, the step
scattering induced Friedel oscillations in the local density of states (LDOS)
exhibit quite unique decaying power law \cite{WangJ,Biswas} as compared to
that in noble metals \cite{Mitsuoka,Davis,Crommie1993,Hasegawa1993} and
graphene \cite{Xue2012}. Furthermore, with the advanced progress of STM
technologies, it is possible to study the optics-analogous properties of the
quantum interference as well as quantum confinement of the conventional
two-dimensional electron gas on noble metals \cite{Avouris1993,Burgi1998}.
Recently, many efforts have also been devoted to investigate the Dirac
electron analogue of optical phenomena, such as negative refraction,
Goos-H\"{a}nchen effect, beam collimation, and Fabry-P\'{e}rot
resonances/interferences, in graphene systems
\cite{And,Buchs,Masir,Campos,AGRAWAL,Rickhaus,Hammer,Oksanen} and TI materials
\cite{Wu2010,Seo,Ferraro,Rizzo}. For instance, Fabry-P\'{e}rot type
conductance oscillation patterns of Dirac fermions in monolayer graphene
\textit{p-n} junctions \cite{Campos,Oksanen} and giant conductance
oscillations in ballistic trilayer graphene Fabry-P\'{e}rot interferometers
\cite{Rickhaus} have been successfully observed in recent quantum transport
experiments. More recently, in contrast to ordinary surface states in common
metals, recent STM experiment performed on TI material Sb(111) surface
suggests that the Dirac fermions transmit the step barriers with a high
probability, and the Fabry-P\'{e}rot resonance of the topological surface
states has been observed \cite{Seo}. Although it is highly challenging in
experiments to confine the massless Dirac fermions in the nanoscale quantum
corrals \cite{FuZG2012} constructed by adatoms or Fabry-P\'{e}rot resonators
formed by straight parallel step defects on TI surface, the latest success has
been achieved on Bi$_{2}$Te$_{3}$(111)-based samples \cite{ChenM}.
\begin{figure}[ptb]
\begin{center}
\includegraphics[width=1.\linewidth]{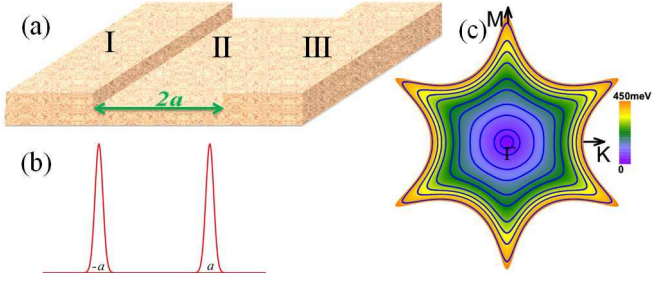}
\end{center}
\caption{ \textbf{Two symmetric steps on a TI surface with strong warping
effect.} (a) Schematics of the symmetric steps and (b) double $\delta
$\texttt{-}type barriers. (c) Constant energy contour of $\varepsilon
_{s}\left(  \boldsymbol{k}\right)  $ for Bi$_{2}$Te$_{3}$ TI material with the
fermi velocity $v_{f}$=240 meV$\cdot$nm, and the warping parameter
$\lambda\mathtt{=}250$ meV$\cdot$nm$^{3}$. }%
\label{fig1}%
\end{figure}

Because of its importance both from basic point of interest and to TI-based
quantum device applications, in the present paper we address this issue by
presenting an attempt at the theoretical evaluation of the Fabry-P\'{e}rot
resonance problem of the massless Dirac electrons on the TI surface in the
presence of double symmetric steps. We show that because of the strong warping
effect in the topological surface states, the electronic LDOS confined in the
step-well running along$\ y$ ($\Gamma\mathtt{-}$M) direction and those in the
step-well running along $x$ ($\Gamma$-K) direction exhibit anisotropic
Fabry-P\'{e}rot resonance images. The resonant transmission properties as well
as the spin orientation of Dirac fermion in both cases are also influenced
remarkably by the warping effect. These findings could be confirmed by STM measurements.

\section*{RESULTS}

We start from the effective Dirac Hamiltonian for the TI surface which is
expressed as \cite{WangJ,FuL, LiuCX}
\begin{equation}
H_{0}\left(  \boldsymbol{k}\right)  \mathtt{=}v_{f}\left(  \sigma_{x}%
k_{y}\mathtt{-}\sigma_{y}k_{x}\right)  \mathtt{+}\frac{\lambda}{2}\left(
k_{+}^{3}\mathtt{+}k_{-}^{3}\right)  \sigma_{z}, \label{h1}%
\end{equation}
$\allowbreak$where $v_{f}$ ($\mathtt{\sim}240$ meV$\cdot$nm) is the Fermi
velocity, $\lambda$ ($\mathtt{\sim}250$ meV$\cdot$nm$^{3}$ for Bi$_{2}$%
Te$_{3}$) is the warping parameter \cite{ZhangT,WangJ,FuL},
$\boldsymbol{\sigma}\mathtt{=}\left\{  \sigma_{x},\sigma_{y},\sigma
_{z}\right\}  $ are Pauli matrices acting on spin space, and $k_{\pm
}\mathtt{=}k_{x}\mathtt{\pm}ik_{y}$. For simplicity, here we ignore the
spin-independent $k^{2}/2m$ term in the effective Hamiltonian since it just
results in particle-hole asymmetry but affects the shape of Fermi surface
little. Therefore, the anisotropic Fabry-P\'{e}rot resonant states discussed
following will not be affected when the particle-hole asymmetry is absent. The
eigenfunction of $H_{0}$ is given by $\psi_{s}\left(  \boldsymbol{k}%
,\boldsymbol{r}\right)  \mathtt{=}\left(  \phi_{1,s},si\phi_{2,s}e^{i\theta
}\right)  ^{\text{T}}e^{i\boldsymbol{k}\cdot\boldsymbol{r}}$, where
$e^{i\theta}\mathtt{=}\frac{k_{x}+ik_{y}}{k}$, and $s\mathtt{=}\mathtt{\pm}1$
corresponds to the upper and lower band dispersion $\varepsilon_{s}\left(
\boldsymbol{k}\right)  \mathtt{=}s\sqrt{\left(  v_{f}k\right)  ^{2}%
\mathtt{+}\lambda^{2}k^{6}\cos^{2}\left(  3\theta\right)  }$, which is
schematically shown in Fig. 1(c). Here, $\phi_{1,s}\mathtt{=}\mathtt{-}%
\frac{k^{\prime}}{\sqrt{2d\left(  d-sd_{3}\right)  }}$, $\phi_{2,s}%
\mathtt{=}\sqrt{\frac{d-sd_{3}}{2d}}$, where $\boldsymbol{d}\mathtt{=}\left\{
d_{1},d_{2},d_{3}\right\}  \mathtt{=}\{k_{y}^{^{\prime}},\mathtt{-}%
k_{x}^{^{\prime}},\left(  k_{+}^{\prime3}\mathtt{+}k_{-}^{\prime3}\right)
/2\}$ with $k^{\prime}\mathtt{=}k\sqrt{\lambda/v_{f}}\mathtt{=}kl$.

Usually, the straight and parallel steps could be naturally formed in the
progress of molecular-beam epitaxy growth, which will result in scattering of
quasiparticles on the surface of the sample. Hereafter we consider the problem
of massless Dirac electrons confined in a pair of symmetric steps apart $2a$
along the $x$ or $y$ direction on the surface of a three-dimensional TI, which
is schematically shown in Fig. 1(a). The step-edge potential is also
illustrated in Fig. 1(b), which is assumed as one-dimensional $\delta
\mathtt{-}$type.

\begin{figure}[ptb]
\begin{center}
\includegraphics[width=1.\linewidth]{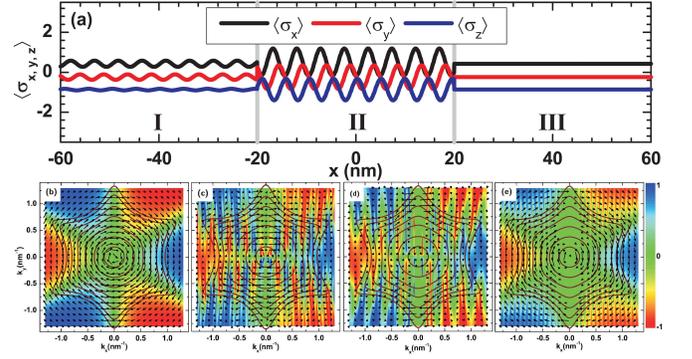}
\end{center}
\caption{ \textbf{Spin orientations of Dirac fermions on the surface of TI.}
(a) The spatial distribution of spin orientation for Dirac fermions scattered
from the double steps, where the incident angle $\theta=\pi/3$. (b) Spin
orientation of free Dirac fermions. (c-e) Spin orientation of scattered Dirac
fermions at Region I $x=-1.5a$ (c), region II $x=0.5a$ (d), and region III (e)
on a TI surface with double steps located at $\pm a$. In (b-e) the arrows
(color backgrounds) denote the $xy-$plane component $\langle\sigma
_{\shortparallel}\rangle$ ($z$ component $\langle\sigma_{z}\rangle$), and the
wine curves denote the constant energy contours. Parameters are chosen as
$a$=20 nm, $s=+1$, $v_{f}$=240 meV$\cdot$nm, warping effect $\lambda$=250
meV$\cdot$nm$^{3}$, and $u_{0}$=$\frac{Vh}{2v_{f}}$=$1.5$. }%
\label{spin}%
\end{figure}

\textbf{Steps along the }$y$\textbf{\ direction.} In what follows, let us
consider the double nanoscale parallel steps along the $y$ direction on TI
surface with the scattering potential described as $\delta\mathtt{-}$type
potential barriers \cite{ZhangDg,AnJ,Mitsuoka,Davis,Crommie1993}
\begin{equation}
U\left(  x\right)  \mathtt{=}Vh\left[  \delta\left(  x\mathtt{+}a\right)
\mathtt{+}\delta\left(  x\mathtt{-}a\right)  \right]  , \label{poten}%
\end{equation}
where $h$ is the hight of step and the step edges are located at
$x_{0}\mathtt{=}\mathtt{\pm}a$. At this stage, we would like to point out that
it is easy, on one side, to treat the boundary conditions for a pair of
$\delta\mathtt{-}$type potential barriers, while for other types of barriers
(such as rectangular, semielliptic, and Gaussian profile barriers) the
eigenequations of the boundary condition become too tedious and complicated to
obtain the analytical results. On the other side, the key physical properties
in the step scattering will not be lost by using $\delta\mathtt{-}$type
barriers herein. Therefore, in this work we choose to employ barrier model
(\ref{poten}) for simplicity and clarification.

In the scattering process, we suppose an incident electron plane wave from one
side [region I in Fig. 1(a)] with momentum $\boldsymbol{k}$ and energy
$\varepsilon_{s}$ will be reflected back to the same side or transmits into
the step-well (region II), and then the transmitted part will be reflected
between the both steps or transmits further into the other side (region III).
Thus, the wave functions could be written as
\begin{equation}
\left\{
\begin{array}
[c]{l}%
\psi_{I}\left(  \boldsymbol{r}\right)  \mathtt{=}\psi_{s}\left(
\boldsymbol{k},\boldsymbol{r}\right)  \mathtt{+}\mathcal{R}\psi_{s}\left(
\boldsymbol{k}_{f},\boldsymbol{r}\right)  ,\text{ \ \ }\left(  x\mathtt{<}%
-a\right) \\
\psi_{II}\left(  \boldsymbol{r}\right)  \mathtt{=}\mathcal{A}\psi_{s}\left(
\boldsymbol{k},\boldsymbol{r}\right)  \mathtt{+}\mathcal{B}\psi_{s}\left(
\boldsymbol{k}_{f},\boldsymbol{r}\right)  ,\text{ \ }\left(  \left\vert
x\right\vert \mathtt{<}a\right)  \text{\ }\\
\psi_{III}\left(  \boldsymbol{r}\right)  \mathtt{=}\mathcal{T}\psi_{s}\left(
\boldsymbol{k},\boldsymbol{r}\right)  ,\text{ \ \ }\left(  x\mathtt{>}%
a\right)
\end{array}
\right.  . \label{wf}%
\end{equation}
Here, we have assumed that $k_{y}$ is a good quantum number and the energy is
conserved in the electron scattering process, naively, the reflected and
transmitted waves will be characterized by $\boldsymbol{k}_{f}\mathtt{\equiv
}(-k_{x},k_{y})\mathtt{=}(k,\theta_{f})$ and $\boldsymbol{k}\mathtt{\equiv
}(k_{x},k_{y})\mathtt{=}(k,\theta)$ with $\theta_{f}\mathtt{=}\pi
\mathtt{-}\theta$. Such assumption is an analogy between electron transport
and light propagation since it could be understood in terms of phenomena like
reflection and transmission \cite{AGRAWAL}. This results in that $\phi
_{1,s}(-k_{x},k_{y})\mathtt{=}\phi_{1,-s}(k_{x},k_{y})$ and $\phi_{2,s}%
(-k_{x},k_{y})\mathtt{=}\phi_{2,-s}(k_{x},k_{y})$. $\mathcal{R}$ and
$\mathcal{T}$ denote the reflection amplitude and tunneling amplitude,
respectively. By considering the boundary conditions of wave functions at the
step edges $x_{0}\mathtt{=}\mathtt{\pm}a$, we can obtain the amplitudes
$\mathcal{A}$, $\mathcal{B}$, $\mathcal{R}$, and $\mathcal{T}$, seeing the
details of the derivation in the section of METHODS.

\begin{figure}[ptb]
\begin{center}
\includegraphics[width=1.\linewidth]{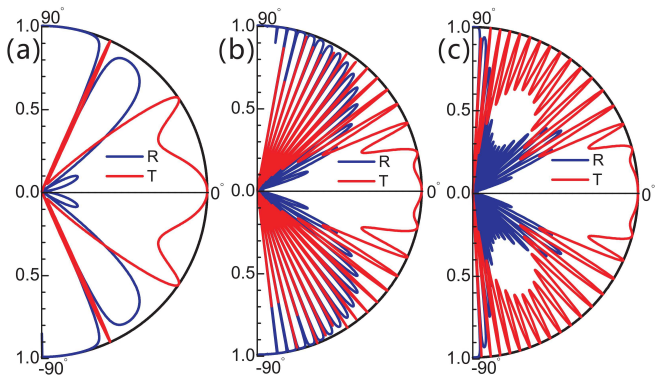}
\end{center}
\caption{ \textbf{Transmittance (red) and reflectance (blue) strengths of
Dirac fermions on the surface of TI with two symmetric steps grown along the
}$y$\textbf{ direction.} (a-b) Transmittance $T$ and reflectance $R$ vs. the
incident angle $\theta$ without warping effect. (c) $T$ and $R$ with warping
effect. The momentum is chosen as $k$=$0.2$ nm$^{-1}$ in (a), and $k$=$1.3$
nm$^{-1}$ in (b) and (c). Other parameters are the same as those in Fig. 2. }%
\label{fig2}%
\end{figure}It is also easy by using Eq. (\ref{wf}) to get the analytical
expressions of spin orientation $\left\langle \boldsymbol{\sigma}\right\rangle
=\left\langle \psi\left(  \boldsymbol{r}\right)  \right\vert
\boldsymbol{\sigma}\left\vert \psi\left(  \boldsymbol{r}\right)  \right\rangle
$ of Dirac fermion on TI surface in the presence of Fabry-P\'{e}rot resonator,
which are shown in the part of METHODS. Because of the scattering and
interference induced by the double steps, the spin texture of the bands
becomes spatially dependent, and is modified via the amplitudes $\mathcal{A}$,
$\mathcal{B}$, $\mathcal{R}$, and $\mathcal{T}$, which could be clearly found
from Eqs. (14\texttt{-}16). As an example, in Fig. 2(a) we exhibit the spatial
distribution of spin orientation for a scattered Dirac fermion with incident
angle $\theta=\pi/3$, where the black line for $\left\langle \sigma
_{x}\right\rangle $, red line for $\left\langle \sigma_{y}\right\rangle $, and
blue line for $\left\langle \sigma_{z}\right\rangle $. It is obvious from Fig.
2(a) that $\left\langle \sigma_{x,y,z}\right\rangle _{x<-a}$ and $\left\langle
\sigma_{x,y,z}\right\rangle _{\left\vert x\right\vert <a}$ in incoming and
step-well regions [\textit{i.e.}, region I and II shown in Fig. 2(a)]
periodically oscillate with space coordinate $x$ while $\left\langle
\sigma_{x,y,z}\right\rangle _{x>a}$ in out-coming region [region III in Fig.
2(c)] are independent on $x$, which are consistent with the analytical results
in Eqs. (14\texttt{-}16). Furthermore, one can also observe from Fig. 2(a)
that the spin orientations of $\left\langle \sigma_{x,y,z}\right\rangle $ are
discontinuous at the steps $x_{0}=\pm a$ indicating sudden change of the
direction of spin, which is similar to the case of a single step (not shown
here for briefness). The numerical results of spin orientation corresponding
to Eqs. (14\texttt{-}16) as a function of the direction of in-plane
wave-vector are shown in Figs. 2(c-e), and for comparison, the spin
orientation for the free Dirac fermion on a clean surface of TI $\left\langle
\boldsymbol{\sigma}\right\rangle =\frac{s}{d}(k^{\prime}\sin\theta,-k^{\prime
}\cos\theta,d_{3})$ is also shown in Fig. 2(b). In Figs. 2(b-e), the black
arrows denote the $xy-$plane component $\langle\sigma_{\shortparallel}%
\rangle=\langle\sigma_{x}\rangle+i\langle\sigma_{y}\rangle$, the color
background denotes the $z$ component $\langle\sigma_{z}\rangle$, and the wine
curves denote the contours of constant energy. We can see from Fig. 2(b) that
for the free Dirac fermion, the $xy\mathtt{-}$plane component $\langle
\sigma_{\shortparallel}\rangle$ rotates clockwise ($s=+1$), and the threefold
symmetry of $\langle\sigma_{z}\rangle$ becomes obvious with increasing the
Fermi energy due to the warping effect. Comparing with the case of free Dirac
fermion [Fig. 2(b)], due to the quantum interference, the spin orientation of
the scattered Dirac fermion located in regions I and II is affected so strong
that, on one hand not only the amplitude of $\langle\sigma_{\shortparallel
}\rangle$ is changed remarkably but also the clockwise rotation of
$\langle\sigma_{\shortparallel}\rangle$ is lost; on the other hand the
threefold symmetry of $\left\langle \sigma_{z}\right\rangle $ is broken, see
Figs. 2(c) and 2(d). While for the Dirac fermion transmitted into region III,
there are two key points need to be noticed that: i) both the property of
clockwise rotation of $\langle\sigma_{\shortparallel}\rangle$ and the
threefold symmetry of $\left\langle \sigma_{z}\right\rangle $ are kept; ii)
the amplitude of $\left\langle \sigma_{x,y,z}\right\rangle _{x>a}$ is
proportional to the transmittance $\left\vert \mathcal{T}\right\vert ^{2}$
[Eq. (14)], and thus that $\left\langle \sigma
_{z}\right\rangle $ exhibits sharp peaks in Fig. 2(e) which is corresponding
to resonant tunneling (seeing the following discussions). Moreover, in the
absence of warping effect of surface bands (\textit{i.e.},\textit{ }$d_{3}=0$
and $k^{\prime}=d$), the spin orientation $\left\langle \boldsymbol{\sigma
}\right\rangle $ could be further simplified (see the formulae below Eq.
(\ref{ss1}) in the METHODS section). Especially, we find in this case that: i)
$\left\langle \sigma_{y}\right\rangle $ will not oscillate in real-space since
it becomes independent on $x$, and ii) the spin of Dirac fermion in region III
is still locked in the surface plane since $\left\langle \sigma_{z}%
\right\rangle _{x>a}=0$. The numerical results for the case without warping
effect is not shown herein for briefness.

\begin{figure}[ptb]
\begin{center}
\includegraphics[width=1.\linewidth]{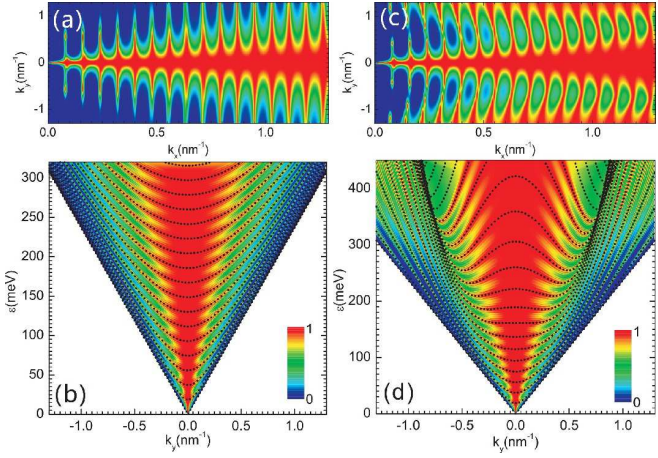}
\end{center}
\caption{\textbf{Resonance transmission of Dirac fermions on the surface of TI
with two symmetric steps grown along the }$y$\textbf{ direction.} (a-b)
Transmittance contour plot versus $k_{x}$ and $k_{y}$, and versus $k_{y}$ and
$\varepsilon$ for the case without warping effect. (c-d) Transmittance for the
case with warping effect. The black dotted curves in (b) and (d) are for the
resonance energy $\varepsilon_{n,s}$. The parameters are the same as those in
Fig. 2. }%
\label{fig3}%
\end{figure}

Now let us turn to discuss the transmission property of Dirac fermions in the
Fabry-P\'{e}rot resonator on TI surface, which is also irradiative for the
explanation of the spin orientation discussed above. Differing from the single
step case \cite{ZhangDg, AnJ,Rakyta}, there will appear resonant tunneling in
two-step or multi-step scattering. We could analytically determine from Eq.
(\ref{ss1}), that the resonance transmission satisfies the condition of
$k_{x}^{n}\mathtt{=}\frac{n\pi}{2a}$ with $n$ an integer number. In other
words, for the incoming state with wave vector $\boldsymbol{k}\mathtt{=}%
(k,\theta)$, the resonance transmission occurs when the incident angle
satisfies the relationship that $\cos\theta$=$\frac{1}{s\sqrt{1+k_{y}%
^{2}/\left(  \frac{n\pi}{2a}\right)  ^{2}}}$, and the corresponding resonance
level is expressed as $\varepsilon_{n,s}\left(  k_{x}^{n},k_{y}\right)
=sv_{f}\sqrt{\left(  \frac{n\pi}{2a}\right)  ^{2}+k_{y}^{2}}$ in the absence
of warping effect in the topological surface state. It is easy to check that
the reflection coefficient and tunneling amplitudes satisfy the relation of
$\left\vert \mathcal{R}\right\vert ^{2}$\texttt{+}$\left\vert \mathcal{T}%
\right\vert ^{2}\mathtt{\equiv}R\mathtt{+}T\mathtt{=}1$. Furthermore, we
should point out that if the plane wave incident from the right side of steps,
\textit{i.e.}, from region III, one could also obtain the similar expressions
of $\mathcal{R}^{\prime}$ and $\mathcal{T}^{\prime}$, and $\left\vert
\mathcal{R}^{\prime}\right\vert ^{2}\mathtt{+}\left\vert \mathcal{T}%
\right\vert ^{2}\mathtt{=}\left\vert \mathcal{R}\right\vert ^{2}%
\mathtt{+}\left\vert \mathcal{T}^{\prime}\right\vert ^{2}\mathtt{=}1$ is
satisfied, which is the same as that of a single step case. Similar to the
case of the optical Fabry-P\'{e}rot resonator, $\mathcal{R}$ $\mathtt{and}$
$\mathcal{T}$ are also determined by the distance between the two mirrors and
the incident angle of light, and if the absorption is neglected $\left\vert
\mathcal{R}\right\vert ^{2}$\texttt{+}$\left\vert \mathcal{T}\right\vert
^{2}\mathtt{=}1$ is required because of the conservation of energy. However,
differing from the optical case, $\mathcal{R}$ $\mathtt{and}$ $\mathcal{T}$ of
Dirac fermions in the present system are strongly dependent on the nature of
bands (warping effect) and the height of barriers seeing the following
discussions and formulae in the METHODS section.

\begin{figure}[ptb]
\begin{center}
\includegraphics[width=1.\linewidth]{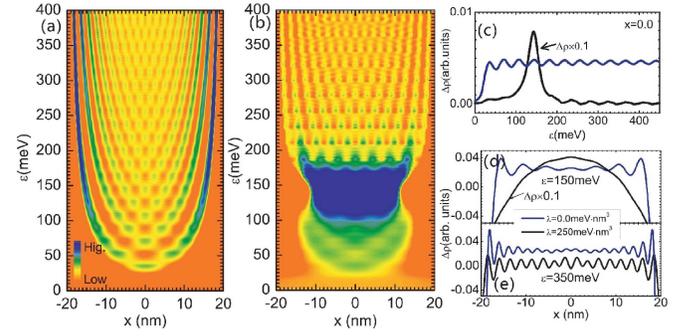}
\end{center}
\caption{\textbf{The change in the LDOS of Dirac electrons confined in the
double symmetric steps apart along the }$y$\textbf{ direction on TI surface.}
(a) The change in the LDOS $\Delta\rho_{II}\left(  x,\varepsilon\right)  $
confined in the double symmetric $\delta-$type steps apart $2a=40$ nm on TI
surface without warping effect. (b) $\Delta\rho_{II}\left(  x,\omega\right)  $
confined in the steps on TI surface with warping effect. (c) Bias dependent
LDOS spectra at the center of step-well $x\mathtt{=}0$. (d-e) Spatially
dependent LDOS spectra within finite bias $\varepsilon\mathtt{=}150$ meV and
$\varepsilon\mathtt{=}350$ meV, respectively. The blue (black) curves are for
the case of without (with) warping effect. All the other parameters are the
same as those in Fig. 2. }%
\label{figLDOS}%
\end{figure}

The typical results of reflectance and transmission strengths as functions of
incident angle $\theta\mathtt{\ }$(the angle of incident wave vector
$\boldsymbol{k}$ to the normal direction of steps) are shown in Fig. 3. One
can find the resonant tunneling $T\mathtt{=}1$ (correspondingly, the
backscattering is forbidden) at normal incidence ($\theta\mathtt{=}0$) and
other special values of $\theta$ that satisfy the resonance conditions, see
the red curves in Fig. 3. Another interesting phenomenon should be noticed
that although the spin orientation of Dirac fermion rotates in the step-well
region due to the quantum interference, when the resonant tunneling occurs its
orientation in the out-coming region will turn back to that in the incoming
region, \textit{i.e.}, $\left\langle \boldsymbol{\sigma}\right\rangle
_{x>a}=\left\langle \boldsymbol{\sigma}\right\rangle _{x<a}=\frac{s}%
{d}(k^{\prime}\sin\theta,-k^{\prime}\cos\theta,d_{3})$ when $T\mathtt{=}1$ and
$R\mathtt{=}0$, seeing Eqs. (14) and (16). In other words, the transmitted
Dirac fermion points along the same direction of the incident fermionic spin
when it resonantly tunnels through single- or multiple-step. It can also be
found from Fig. 3 that with increasing the energy of incident electron, the
resonant tunneling becomes more remarkable. By a comparison of Fig. 3(b) and
Fig. 3(c), one can see that the curves of reflectance and transmittance
strengths are reformed due to the influence of strong warping effect, but the
incident angles corresponding to the resonant tunneling are unaltered.

In order to determine the discrete resonance energy levels confined in the
step-well, we plot the transmittance strength versus the wave vector
components $k_{x}$ and $k_{y}$ without [Fig. 4(a)] and with [Fig. 4(c)]
warping effect. Correspondingly, the colored background in Figs. 4(b) and 4(d)
are the transmittance strength versus the energy $\varepsilon$ of electron and
the wave vector component $k_{y}$. With increasing energy, it is clear that
the warping effect becomes strong and obvious. We can observe the resonant
tunneling peaks corresponding to the quantized $k_{x}$, and from these
numerical results of $T$ the resonance condition can be also concluded as
$k_{x}^{n}\mathtt{=}\frac{n\pi}{2a}$, which is consistent with the analytical
results obtained from the expression for $\mathcal{T}$ in the absence of
warping term [see Eq. (\ref{ss1})]. Substituting this $k_{x}^{n}$ back into
the band dispersion $\varepsilon_{s}\left(  \boldsymbol{k}\right)  $, one can
obtain the discrete resonance energy levels $\varepsilon_{n,s}\left(
k_{x}^{n},k_{y}\right)  $ confined in the step-well, which are also shown in
Figs. 4(b) and 4(d) by the black dotted curves. It is clear that the resonance
$\varepsilon_{n,s}\left(  k_{x}^{n},k_{y}\right)  $ are perfectly consistent
with the resonant tunneling peaks.

\begin{figure}[ptb]
\begin{center}
\includegraphics[width=1.\linewidth]{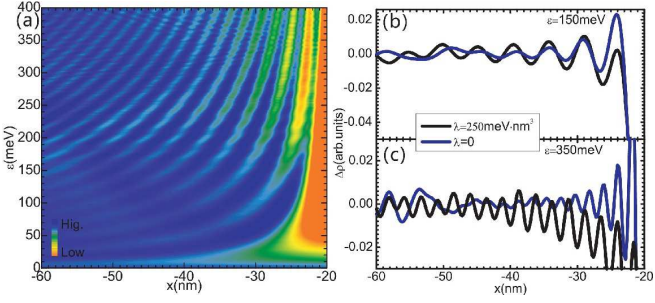}
\end{center}
\caption{\textbf{The change in the LDOS outside of the double steps along the
}$y$\textbf{ direction on TI surface with warping effect. }(a) The change in
the LDOS $\Delta\rho_{I}\left(  x,\varepsilon\right)  $ versus $x$ and
$\varepsilon$. (b) and (c) spatially dependent $\Delta\rho_{I}\left(
x,\varepsilon\right)  $ spectra within finite bias $\varepsilon\mathtt{=}150$
meV and $\varepsilon\mathtt{=}350$ meV, respectively. The blue (black) curves
are for the case of without (with) warping effect. }%
\label{figLDOS-ref}%
\end{figure}

After obtaining the resonance energy levels, one can easily get the LDOS
between the double steps, which is expressed as
\begin{equation}
\rho_{II}\left(  \boldsymbol{r},\varepsilon\right)  \mathtt{=}\sum_{k_{x}%
^{n},s}\int\frac{dk_{y}}{2\pi}\left\vert \psi_{II}\left(  s,k_{x}^{n}%
,k_{y},\boldsymbol{r}\right)  \right\vert ^{2}\delta\left(  \varepsilon
\mathtt{-}\varepsilon_{n,s}\right)  . \label{DOS}%
\end{equation}
In order to observe the Fabry-P\'{e}rot resonant states more clearly, we
subtract the contributions of the averaged background $\rho_{II}^{0}$ and only
consider the change in the LDOS $\Delta\rho_{II}\left(  \boldsymbol{r}%
,\varepsilon\right)  $ due to the step scattering in following calculations.

Figure 5 shows the energy and position dependence of the LDOS confined in a
symmetric step-well along the $y$ ($\Gamma\mathtt{-}$M) direction on TI
surface. One can clearly see the obvious quantum confinement, and the LDOS
images are symmetric because our Fabry-P\'{e}rot step-well model is symmetric
in the calculations herein. The number of resonant peaks increases with
increasing energy, which is consistent with the resonance properties of
transmittance shown above. In our calculations, the ground state\ confined in
the Fabry-P\'{e}rot resonator is located at $40$ meV, while the first excited
state\ with one node lies at $55$ meV. When comparing Fig. 5(a) with Fig.
5(b), it is surprising to find that there exists a strong enhancement in the
LDOS in the energy region from $120$ meV to $200$ meV when the warping effect
is taken into account in the calculations ($\lambda\mathtt{\neq}0$), while at
other energy regions the resonance modes are similar to the case in the
absence of warping effect ($\lambda\mathtt{=}0$). The differences in peak
intensity and resonance period could be more clearly observed from the curves
in Figs. 5(c)-(e). In addition, we have also checked other step-wells with
different widthes, the enhancement of the LDOS around the energy region of
$120\mathtt{\sim}200$ meV is still remarkable (not shown). This enhancement is
closely related to the occurrence of the flat-band crossover when the $k_{y}%
$-linear positive term, which is the dominate term at low energy, is
superposed onto the $k_{y}$-square negative term which is dominate at high
energy [see Fig. 3(d)]. This bunches the electronic states and therefore
results in the enhancement of the LDOS in the observed energy region. One
could also qualitatively understand the enhancement of LDOS in the energy
region of 120$\sim$200 meV from the view point of the approximation of
stationary phase points (SPPs) \cite{WangJ}. The LDOS is dominated by the SPPs
[$(k_{x},k_{y})$ and $(-k_{x},k_{y})$] along the direction of $\Gamma
\mathtt{-}$K on the constant energy contour since most of the contribution to
the integral in Eq. (\ref{DOS}) to be from the vicinity of these points.
Therefore, the integral can be broken up into regions between stationary phase
points that $\Delta\rho_{II}\left(  x,\varepsilon\right)  \mathtt{\approx
}\operatorname{Re}\left[  \sum_{k_{x}^{n}\in SPPs}%
{\displaystyle\int\nolimits_{e}}
AB^{\ast}\frac{k^{\prime}\left(  1-e^{2i\theta}\right)  }{2d\pi}e^{2ik_{x}%
^{n}x}dk_{y}\right]  $, where $%
{\displaystyle\int\nolimits_{e}}
$ means the integration of $k_{y}$ are around SPPs. Taking into account the
warping effect, the separation between the resonance levels at energy region
of 120$\sim$200 meV ($n=5,6,7,8$) is compressed, seeing the black dotted
curves in Fig. 4(d), and thus that the LDOS will be enhanced. While with
further increasing of energy, the separation between resonance levels becomes
larger and larger, which results in the decrease of LDOS [see Fig. 5(b) and
the black curves in Fig. 5(c) and Fig. 5(d)]. Additional $\boldsymbol{q}%
$-vectors ($\boldsymbol{q=k}_{f}-\boldsymbol{k}_{i}$) apart from the SPPs will
blur the LDOS since the rapidly varying phase factor $e^{i\left(  k_{x}%
^{i}-k_{x}^{f}\right)  x}$ in the integral of Eq. (\ref{DOS}), which has
little contribution to the total LDOS. \begin{figure}[ptb]
\begin{center}
\includegraphics[width=1.\linewidth]{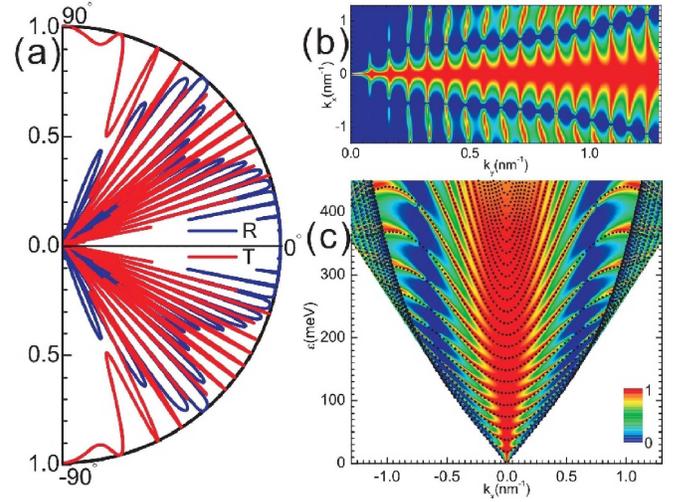}
\end{center}
\caption{\textbf{Transmittance and reflectance strengths of Dirac fermions on
the surface of TI with two symmetric steps grown along the }$x$\textbf{
direction. }(a) Transmittance $T$ and reflectance $R$ versus the incident
angle $\theta$. (b) Transmittance contourplot vs $k_{x}$ and $k_{y}$, and (c)
transmittance contourplot vs $k_{x}$ and $\varepsilon$ (c). The warping effect
is taken into account in calculations. $k$=$1.3$ nm$^{-1}$ in (a). The black
dotted curves in (c) are for the resonance energy $\varepsilon_{n,k}$. }%
\label{fig-y}%
\end{figure}

The numerical results of LDOS outside the step-well $\Delta\rho_{I}\left(
\boldsymbol{r},\varepsilon\right)  $ are plotted in Fig. 6. We can see from
Fig. 6(a) that resonant tunneling of the Dirac fermions is obvious in the
contour plot of energy-resolved modulations of the LDOS, which is consistent
with the above discussions of the transmittance. Different from the single
step case, the quantum interferences due to the multi-step scattering is clear
in the present Fabry-P\'{e}rot resonator, and the warping effect is also
remarkable [comparing the black curves with the blue curves in Figs. 6(b) and
6(c)] with the increase of energy. In addition, the character of the
Friedel-type decaying oscillations in the present case is similar to the
single step case. Our theoretical simulations are qualitatively consistent
with the STM experimental observations of the Fabry--P\'{e}rot resonance of
the topological surface states on the Sb(111) surface \cite{Seo}, and the
theoretical methods used in this work should be useful for the explanations of
the related STM experiments. One should also notice that because $k_{y}$ keeps
unchanged in the scattering process, the LDOS oscillates only in the
perpendicular direction of steps, which is the same as the spin orientation.

\textbf{Steps along the x direction.}\textit{ }Now we switch gears and
consider the case of steps along the $x$ ($\Gamma\mathtt{-}$K) direction. Due
to the hexagonal warping term in the Hamiltonian (\ref{h1}), the $x$
($\Gamma\mathtt{-}$K) and $y$ ($\Gamma\mathtt{-}$M) directions are
inequivalent, which could be seen from the constant energy contour shown in
Fig. 1(c). Moreover, the warping term in Hamiltonian (\ref{h1}) is represented
as $i\lambda\left(  \partial_{x}^{3}\mathtt{-}\partial_{x}\partial_{y}%
^{2}\right)  \sigma_{z}$ in the real space, therefore, it is naturally
expected that different orientation of steps leads to different
Fabry-P\'{e}rot resonant features in the LDOS and in the
reflection/transmittance properties. For the steps along the $x$
($\Gamma\mathtt{-}$K) direction, the wave function in each region should be
written as
\begin{equation}
\left\{
\begin{array}
[c]{l}%
\psi_{I}\left(  \boldsymbol{r}\right)  \mathtt{=}\psi_{s}\left(
\boldsymbol{k},\boldsymbol{r}\right)  \mathtt{+}\overline{\mathcal{R}}\psi
_{s}\left(  \boldsymbol{k}_{f},\boldsymbol{r}\right)  ,\text{ \ \ }\left(
y\mathtt{<}-a\right) \\
\psi_{II}\left(  \boldsymbol{r}\right)  \mathtt{=}\overline{\mathcal{A}}%
\psi_{s}\left(  \boldsymbol{k},\boldsymbol{r}\right)  \mathtt{+}%
\overline{\mathcal{B}}\psi_{s}\left(  \boldsymbol{k}_{f},\boldsymbol{r}%
\right)  ,\text{ \ \ }\left(  \left\vert y\right\vert \mathtt{<}a\right) \\
\psi_{III}\left(  \boldsymbol{r}\right)  \mathtt{=}\overline{\mathcal{T}}%
\psi_{s}\left(  \boldsymbol{k},\boldsymbol{r}\right)  ,\text{ \ \ }\left(
y\mathtt{>}a\right)
\end{array}
\right.  . \label{wfy}%
\end{equation}
In this case, the reflected wave vectors will be characterized by
$\boldsymbol{k}_{f}\mathtt{\equiv}(k_{x},-k_{y})\mathtt{=}(k,-\theta)$, which
results in $\phi_{1,s}(k_{x},-k_{y})\mathtt{=}\phi_{1,s}(k_{x},k_{y})$ and
$\phi_{2,s}(k_{x},-k_{y})\mathtt{=}\phi_{2,s}(k_{x},k_{y})$. Following the
same method used above to consider the boundary conditions at $y_{0}%
\mathtt{=}\mathtt{\pm}a$, one could also easily get the coefficients
$\overline{\mathcal{A}}$, $\overline{\mathcal{B}}$, $\overline{\mathcal{R}}$,
and $\overline{\mathcal{T}}$ [see Eq. (\ref{14}) shown in the section of METHODS].

The reflectance and transmittance strengths for this case are shown in Fig. 7.
Similar to the case of steps along $y$ ($\Gamma\mathtt{-}$M) direction, we can
observe the resonant tunneling herein, and the resonance condition could be
concluded as $k_{y}^{n}\mathtt{=}\frac{n\pi}{2a}$. However, the transmittance
mirages and the quantized resonance energy $\varepsilon_{n,s}\left(
k_{x},k_{y}^{n}\right)  $ are different from those for the steps along $y$
($\Gamma\mathtt{-}$M) direction, which could be clearly observed by comparing
Fig. 7 with Fig. 4. Here, the quantized energy dispersion is a result of the
superposition between the $k_{x}$-linear positive term and the $k_{x}$-cubic
positive term. Therefore, the aforementioned flat-band crossover cannot occur
in the present case, as verified in Fig. 7. Notice that $\theta$ in Fig. 7 is
defined in the same way as in Fig. 4, \textit{i.e.}, the angle of incident
wave vector to the positive $x$ axis. Besides, one could derive the spin
orientation in the present case, and the anisotropy could also be found in the
spin orientation since it is related to the coefficients $\overline
{\mathcal{A}}$, $\overline{\mathcal{B}}$, $\overline{\mathcal{R}}$, and
$\overline{\mathcal{T}}$ (not shown for briefness).

\begin{figure}[ptb]
\begin{center}
\includegraphics[width=1.\linewidth]{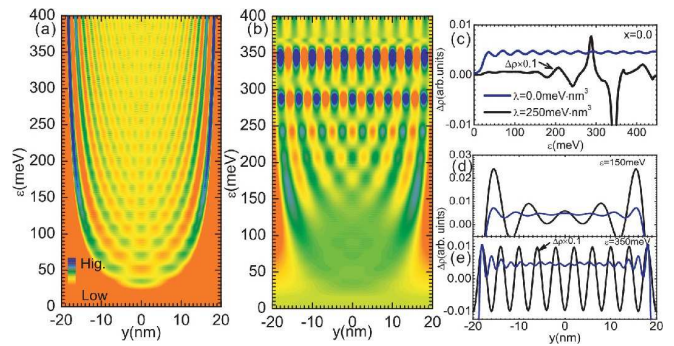}
\end{center}
\par
.\caption{$\Delta\rho_{II}\left(  y,\omega\right)  $\textbf{ of Dirac
electrons confined in the double symmetric }$\delta-$\textbf{type steps apart
}$2a=40$\textbf{ nm along the }$x$\textbf{ direction on TI surface. }The
arrangement is the same as in Fig. 5. }%
\label{figLDOSy}%
\end{figure}The LDOS results for this case are illustrated in Fig. 8. When the
warping effect is taken into account, the Fabry-P\'{e}rot-like resonance of
LDOS confined in the step-well along the $x$ ($\Gamma\mathtt{-}$K) direction
is quite different from the case of steps along the $y$ ($\Gamma\mathtt{-}$M)
direction, which indicates the anisotropic step-scattering character on the TI
surface. In this case, the LDOS is dominated by the SPPs along $\Gamma
\mathtt{-}$M direction on the constant energy contour. Especially, the
Fabry-P\'{e}rot resonance is enhanced [seeing Fig. 8(b)] around
$300\mathtt{\sim}350$ meV in this case since separation between the resonance
levels at this energy region is compressed [seeing the black dotted curves in
Fig. 7(c)] due to the warping effect. These interesting\ anisotropic
Fabry-P\'{e}rot resonance states confined within nano-steps on the TI surface
can not be observed on conventional metal surface or graphene.

\section*{DISCUSSION}

Previous STM experiment \cite{ZhangT} and theory \cite{WangJ} about the
scattering of surface states off single step edges in TI materials Bi$_{2}%
$Se$_{3}$ and Bi$_{2}$Te$_{3}$ in which the warping effect is strong suggested
that the power law decay of the standing waves could be modulated with
increasing the Fermi level due to the strong warping effect. Based on the
above results, we would like to clarify that the experimental relevance of the
anisotropic Fabry-P\'{e}rot-like resonances states on TI surface will also
strongly depend on whether the Fermi level lies close the Dirac point or lies
far from the Dirac point so that transport at Fermi level can get strongly
effect by warping. The modulation of Fermi level could be easily realized at
present experimental conditions by appropriate doping. Chen \textit{et al}.
\cite{Chen} experimentally suggested that the Fermi level of Bi$_{2}$Te$_{3}$
can be tuned to intersect only the surface states with strong warping effect,
and a full energy gap for the bulk states was clearly observed in the
angle-resolved photoemission spectroscopy experiment. This indicates that as
the Fermi level goes away from the Dirac point, although the angle between
spin and momentum is no longer locked in the surface plane since an additional
component of out of plane orientation of spin is raised by the warping effect,
the bulk disorder may play a weak role in the scattering process of
topological surface states. Thereby, we have not considered the role of the
bulk disorder in our calculations.

Furthermore, on one side, similar to the case of scattering of surface states
caused by single-step defect on the surface of TI \cite{ZhangDg, AnJ,Rakyta},
anisotropic transmittance and decay of LDOS outside of the nano-step well has
also been observed in our studies due to the strong warping effect; On the
other side, however, apart from various details of the model, the major
distinctive features of our study are (i) the resonance and interference
behaviors of Dirac fermions transmitting double step barriers as well as the
corresponding resonance energy levels [Figs. 3, 4, 7], and (ii) the
anisotropic Fabry-P\'{e}rot resonant states confined within nano-step wells
grown along different high symmetry directions on TI surface in the presence
of strong warping effect in surface states. Recently, an STM measurement
showed that the Fabry-P\'{e}rot resonance of the topological surface states on
Sb(111) surface with multiple asymmetric steps has been successfully observed
\cite{Seo}. The high-quality Fabry-P\'{e}rot interferences of Dirac fermions
in graphene-based devices have also been achieved in recent quantum transport
experiments \cite{Campos,Rickhaus,Oksanen}. More recently, both the quantum
confinement of massless Dirac fermions in nanoscale quantum corrals surrounded
by Bi-bilayer on Bi$_{2}$Te$_{3}$ surface \cite{ChenM} and the anisotropic
scattering of surface state electrons at a point defect on Bi(111) surface
\cite{Cottin} have been experimentally detected. These experimental
observations indicate that our methods should be helpful for understanding
these experimental findings, and we expect that the anisotropic
Fabry-P\'{e}rot resonance states found in this work could be confirmed in
future STM experiments.

In summary, the Fabry-P\'{e}rot-like resonances of massless Dirac electron
confined within symmetric steps along the $x$ ($\Gamma\mathtt{-}$K) or $y$
($\Gamma\mathtt{-}$M) directions on TI surface have been studied. The
analytical expressions for the spin orientations, reflectance, and
transmission were obtained in the presence of the warping effect. We found
that the spin orientation of Dirac fermion rotates in the step-well region due
to the quantum interference, but the transmitted Dirac fermion points along
the same direction of the incident fermionic spin when it resonantly tunnels
through the step-well. Particullarly, because of the strong warping effect in
the topological surface state, the spin is no longer locked in the surface
plane, and the electronic LDOS image confined in the step-well running along
$\Gamma\mathtt{-}$K direction is remarkably different from that in the
step-well running along $\Gamma\mathtt{-}$M direction. The resonant
transmittance properties as well as the spin orientation of topological
surface states in both cases are also different due to the warping effect.
Furthermore, the formula obtained here can be extended to symmetric/asymmetric
multi-step cases as well as spin polarized cases. Our anisotropic results,
which can not be observed in the conventional metal surface and graphene
system, may be useful for exploring the properties of topological surface
states both for fundamental studies and for evaluating their potential for
device applications.

\section*{METHODS}

We derive an equation of continuation of wave functions at boundaries to get
the coefficients $\mathcal{A}$, $\mathcal{B}$, $\mathcal{R}$, and
$\mathcal{T}$ for the case of the steps grown along the $y$ ($\Gamma
\mathtt{-}$M) direction, which is used to calculate the LDOS in our system.
Within the Eq. (\ref{wf}), one could get the total wavefunction as
\begin{widetext}
\begin{equation}
\psi_{tot}=\psi_{I}\Theta\left(  -a-x\right)  +\psi_{II}\left[  \Theta\left(
a+x\right)  -\Theta\left(  x-a\right)  \right]  +\psi_{III}\Theta\left(
x-a\right)  ,\label{wft}%
\end{equation}
where $\Theta\left(  z\right)  $ is the step function. Now we consider the
property of wavefunction at $x_{0}=\pm a$ respectively. Substituting Eq.
(\ref{wft}) into the Hamiltonian (\ref{h1}) in real space and plus the
$\delta\mathtt{-}$type potential, we have%
\begin{align}
&  \left[  i\lambda\left(  \partial_{x}^{3}-3\partial_{x}\partial_{y}%
^{2}\right)  \sigma_{z}-iv_{f}\partial_{y}\sigma_{x}+iv_{f}\partial_{x}%
\sigma_{y}\right]  \psi_{tot}\left(  x,y\right)  \nonumber\\
&  =\left\{  \varepsilon-Vd\left[  \delta\left(  x\mathtt{+}a\right)
\mathtt{+}\delta\left(  x\mathtt{-}a\right)  \right]  \right\}  \psi
_{tot}\left(  x,y\right)  .\label{w1}%
\end{align}
Integrating both sides of the Eq. (\ref{w1}) near the points $x_{0}=\pm a$,
\begin{align}
&  \lim_{\epsilon\rightarrow0}\int_{\pm a-\epsilon}^{\pm a+\epsilon}dx\left[
i\lambda\left(  \partial_{x}^{3}-3\partial_{x}\partial_{y}^{2}\right)
\sigma_{z}-iv_{f}\partial_{y}\sigma_{x}+iv_{f}\partial_{x}\sigma_{y}\right]
\psi_{tot}\left(  x,y\right)  \nonumber\\
&  =\lim_{\epsilon\rightarrow0}\int_{\pm a-\epsilon}^{\pm a+\epsilon}dx\left[
\varepsilon-Vd\delta\left(  x\mp a\right)  \right]  \psi_{tot}\left(
x,y\right)  ,
\end{align}
we have the following algebraic equations
\begin{equation}
\left\{
\begin{array}
[c]{c}%
a_{0}+a_{1}\mathcal{R}+a_{2}\mathcal{A}+a_{3}\mathcal{B}=0\\
b_{0}+b_{1}\mathcal{R}+b_{2}\mathcal{A}+b_{3}\mathcal{B}=0\\
c_{1}\mathcal{T}+c_{2}\mathcal{A}+c_{3}\mathcal{B}=0\\
f_{1}\mathcal{T}+f_{2}\mathcal{A}+f_{3}\mathcal{B}=0
\end{array}
\right.  ,\label{dsf}%
\end{equation}
where the parameters are expressed as
\begin{equation}%
\begin{array}
[c]{l}%
a_{0}\mathtt{=}\left[  \left(  w\mathtt{+}u\right)  \phi_{1,s}\mathtt{-}%
s\phi_{2,s}e^{i\theta}\right]  e^{-ik_{x}a},\\
a_{1}\mathtt{=}\left[  \left(  w\mathtt{+}u\right)  \phi_{1,-s}\mathtt{+}%
s\phi_{2,-s}e^{-i\theta}\right]  e^{ik_{x}a},\\
a_{2}\mathtt{=}\left[  \left(  u\mathtt{-}w\right)  \phi_{1,s}\mathtt{+}%
s\phi_{2,s}e^{i\theta}\right]  e^{-ik_{x}a},\text{\ }\\
a_{3}\mathtt{=}\left[  \left(  u\mathtt{-}w\right)  \phi_{1,-s}\mathtt{-}%
s\phi_{2,-s}e^{-i\theta}\right]  e^{ik_{x}a},\\
c_{1}\mathtt{=}\left[  \left(  u\mathtt{-}w\right)  \phi_{1,s}\mathtt{+}%
s\phi_{2,s}e^{i\theta}\right]  e^{ik_{x}a},\\
c_{2}\mathtt{=}\left[  \left(  w\mathtt{+}u\right)  \phi_{1,s}\mathtt{-}%
s\phi_{2,s}e^{i\theta}\right]  e^{ik_{x}a},\\
c_{3}\mathtt{=}\left[  \left(  w\mathtt{+}u\right)  \phi_{1,-s}\mathtt{+}%
s\phi_{2,-s}e^{-i\theta}\right]  e^{-ik_{x}a},
\end{array}%
\begin{array}
[c]{l}%
b_{0}\mathtt{=}\left[  \left(  u\mathtt{-}w\right)  s\phi_{2,s}e^{i\theta
}\mathtt{-}\phi_{1,s}\right]  e^{-ik_{x}a},\\
b_{1}\mathtt{=}\left[  \left(  w\mathtt{-}u\right)  s\phi_{2,-s}e^{-i\theta
}\mathtt{-}\phi_{1,-s}\right]  e^{ik_{x}a},\\
b_{2}\mathtt{=}\left[  \left(  w\mathtt{+}u\right)  s\phi_{2,s}e^{i\theta
}\mathtt{+}\phi_{1,s}\right]  e^{-ik_{x}a},\\
b_{3}\mathtt{=}\left[  \mathtt{-}\left(  w\mathtt{+}u\right)  s\phi
_{2,-s}e^{-i\theta}\mathtt{+}\phi_{1,-s}\right]  e^{ik_{x}a},\\
f_{1}\mathtt{=}\left[  \left(  w\mathtt{+}u\right)  s\phi_{2,s}e^{i\theta
}\mathtt{+}\phi_{1,s}\right]  e^{ik_{x}a},\\
f_{2}\mathtt{=}\left[  \left(  u\mathtt{-}w\right)  s\phi_{2,s}e^{i\theta
}\mathtt{-}\phi_{1,s}\right]  e^{ik_{x}a},\\
f_{3}\mathtt{=}\left[  \left(  w\mathtt{-}u\right)  s\phi_{2,-s}e^{-i\theta
}\mathtt{-}\phi_{1,-s}\right]  e^{-ik_{x}a}.
\end{array}
\end{equation}
\end{widetext}
From Eq. (\ref{dsf}), one can get the coefficients $\mathcal{A}$,
$\mathcal{B}$, $\mathcal{R}$, and $\mathcal{T}$. In the presence of the
warping effect, the coefficients are easily written as
\begin{equation}
\mathcal{A}\mathtt{=}\frac{\alpha}{M},\mathcal{B}\mathtt{=}\frac{\beta}%
{M},\mathcal{R}\mathtt{=}\frac{\gamma}{M},\mathcal{T}\mathtt{=}\frac{\tau}{M},
\end{equation}
where $M=\left(  a_{2}b_{1}-a_{1}b_{2}\right)  \left(  c_{3}f_{1}-c_{1}%
f_{3}\right)  -\left(  a_{3}b_{1}-a_{1}b_{3}\right)  \left(  c_{2}f_{1}%
-c_{1}f_{2}\right)  $, $\gamma=\left(  a_{0}b_{3}-a_{3}b_{0}\right)  \left(
c_{1}f_{2}-c_{2}f_{1}\right)  +\left(  a_{2}b_{0}-a_{0}b_{2}\right)  \left(
c_{1}f_{3}-c_{3}f_{1}\right)  $, $\alpha=\left(  a_{0}b_{1}-a_{1}b_{0}\right)
\left(  c_{1}f_{3}-c_{3}f_{1}\right)  $, $\beta=\left(  a_{1}b_{0}-a_{0}%
b_{1}\right)  \left(  c_{1}f_{2}-c_{2}f_{1}\right)  $, and $\tau=\left(
a_{1}b_{0}-a_{0}b_{1}\right)  \left(  c_{2}f_{3}-c_{3}f_{2}\right)  $. Here,
$u\mathtt{=}\frac{Vh}{2iv_{f}}\mathtt{=}\mathtt{-}u_{0}i$ and $w\mathtt{=}%
\left(  k_{x}^{2}\mathtt{-}3k_{y}^{2}\right)  l^{2}$ are dimensionless
variables. We should point out that the properties of Dirac $\delta$-function
that $\lim_{\epsilon\rightarrow0}\int_{a-\epsilon}^{a+\epsilon}dx\delta\left(
x-a\right)  f\left(  x\right)  =\frac{1}{2}\left[  f\left(  a^{+}\right)
+f\left(  a^{-}\right)  \right]  $, and $\int dxf\left(  x\right)
\delta^{\left(  n\right)  }\left(  x\right)  =-\int dx\frac{\partial f\left(
x\right)  }{\partial x}\delta^{\left(  n-1\right)  }\left(  x\right)  $ have
been used in the derivations of Eq. (\ref{dsf}). After obtaining the
wavefunctions, we could easily get the analytical expressions of spin
orientation $\left\langle \boldsymbol{\sigma}\right\rangle =\left\langle
\psi\left(  \boldsymbol{r}\right)  \right\vert \boldsymbol{\sigma}\left\vert
\psi\left(  \boldsymbol{r}\right)  \right\rangle $ of Dirac fermions on TI
surface in the presence of Fabry-P\'{e}rot resonator. In the region I
(incoming region) in Fig. 1(a), we have
\begin{widetext}
\begin{align}
\left\langle \sigma_{x}\right\rangle _{x<-a} &  =\frac{sk^{\prime}}{d}\left(
1+\left\vert \mathcal{R}\right\vert ^{2}\right)  \sin\theta
+2s\operatorname{Im}\left[  \mathcal{R}^{\ast}e^{i\left(  2k_{x}%
x+\theta\right)  }\right]  ,\nonumber\\
\left\langle \sigma_{y}\right\rangle _{x<-a} &  =-\frac{sk^{\prime}}%
{d}\left\vert \mathcal{T}\right\vert ^{2}\cos\theta+2\frac{d_{3}}%
{d}\operatorname{Re}\left[  \mathcal{R}^{\ast}e^{i\left(  2k_{x}%
x+\theta\right)  }\right]  ,\tag{12}\label{y}\\
\left\langle \sigma_{z}\right\rangle _{x<-a} &  =\frac{sd_{3}}{d}\left\vert
\mathcal{T}\right\vert ^{2}+\frac{k^{\prime}}{d}\operatorname{Re}\left[
\mathcal{R}^{\ast}e^{2ik_{x}x}\left(  e^{2i\theta}+1\right)  \right]
.\nonumber
\end{align}
Similarly, in the region II (step-well region), we have
\begin{align}
\left\langle \sigma_{x}\right\rangle _{\left\vert x\right\vert <a} &
=\frac{sk^{\prime}}{d}\left(  \left\vert \mathcal{A}\right\vert ^{2}%
+\left\vert \mathcal{B}\right\vert ^{2}\right)  \sin\theta+2s\operatorname{Im}%
\left[  \mathcal{AB}^{\ast}e^{i\left(  2k_{x}x+\theta\right)  }\right]
,\nonumber\\
\left\langle \sigma_{y}\right\rangle _{\left\vert x\right\vert <a} &
=\frac{sk^{\prime}}{d}\left(  \left\vert \mathcal{B}\right\vert ^{2}%
-\left\vert \mathcal{A}\right\vert ^{2}\right)  \cos\theta+2\frac{d_{3}}%
{d}\operatorname{Re}\left[  \mathcal{AB}^{\ast}e^{i\left(  2k_{x}%
x+\theta\right)  }\right]  ,\tag{13}\label{y2}\\
\left\langle \sigma_{z}\right\rangle _{\left\vert x\right\vert <a} &
=\frac{k^{\prime}}{d}\operatorname{Re}\left[  \mathcal{AB}^{\ast}e^{2ik_{x}%
x}\left(  e^{2i\theta}+1\right)  \right]  \nonumber\\
&  +\frac{sd_{3}}{d}\left\{  \left\vert \mathcal{A}\right\vert ^{2}-\left\vert
\mathcal{B}\right\vert ^{2}-\operatorname{Re}\left[  \mathcal{AB}^{\ast
}e^{2ik_{x}x}\left(  e^{2i\theta}-1\right)  \right]  \right\}  ,\nonumber
\end{align}
\end{widetext}
and in the region III (out-coming region), we have
\begin{align}
\left\langle \sigma_{x}\right\rangle _{x>a} &  =\frac{sk^{\prime}}%
{d}\left\vert \mathcal{T}\right\vert ^{2}\sin\theta,\nonumber\\
\left\langle \sigma_{y}\right\rangle _{x>a} &  =-\frac{sk^{\prime}}%
{d}\left\vert \mathcal{T}\right\vert ^{2}\cos\theta,\tag{14}\label{y3}\\
\left\langle \sigma_{z}\right\rangle _{x>a} &  =\frac{sd_{3}}{d}\left\vert
\mathcal{T}\right\vert ^{2}.\nonumber
\end{align}
In particular, if the warping effect is ignored ($\lambda=0$) in the
derivations, the coefficients should be reduced to the following simpler form
that
\begin{align}
\mathcal{A} &  =\frac{s\left(  1-u^{2}\right)  \cos\theta\left[  2u+s\left(
1+u^{2}\right)  \cos\theta\right]  }{M_{0}},\nonumber\\
\mathcal{B} &  =\frac{se^{i\left(  2k_{x}a-\theta\right)  }u\left(
1-u^{2}\right)  \left(  e^{4i\theta}-1\right)  }{2M_{0}},\nonumber\\
\mathcal{R} &  =\frac{u\left(  e^{2i\theta}-1\right)  \left[  r_{-}%
e^{2ik_{x}a}+r_{+}e^{-2ik_{x}a}\right]  }{M_{0}},\nonumber\\
\mathcal{T} &  =\frac{e^{-2i\theta}\left(  e^{2i\theta}+1\right)  ^{2}\left(
1-u^{2}\right)  ^{2}}{4M_{0}},\tag{15}\label{ss1}%
\end{align}
where $M_{0}=\left[  2u+s\left(  1+u^{2}\right)  \cos\theta\right]
^{2}-2e^{4ik_{x}a}u^{2}\left[  1-\cos\left(  2\theta\right)  \right]  $, and
$r_{\pm}=s\left(  1+u^{2}\right)  \cos\theta\pm2$. Correspondingly, the spin
orientation without considering the warping effect could be simplified, and we
have in region I $\left\langle \sigma_{x}\right\rangle _{x<-a}=s\left(
1+\left\vert \mathcal{R}\right\vert ^{2}\right)  \sin\theta
+2s\operatorname{Im}\left[  \mathcal{R}^{\ast}e^{i\left(  2k_{x}%
x+\theta\right)  }\right]  $, $\left\langle \sigma_{y}\right\rangle
_{x<-a}=-s\left\vert \mathcal{T}\right\vert ^{2}\cos\theta$, and $\left\langle
\sigma_{z}\right\rangle _{x<-a}=\operatorname{Re}\left[  \mathcal{R}^{\ast
}e^{2ik_{x}x}\left(  e^{2i\theta}+1\right)  \right]  $, in region II
$\left\langle \sigma_{x}\right\rangle _{\left\vert x\right\vert <a}=s\left(
\left\vert \mathcal{A}\right\vert ^{2}+\left\vert \mathcal{B}\right\vert
^{2}\right)  \sin\theta+2s\operatorname{Im}\left[  \mathcal{AB}^{\ast
}e^{i\left(  2k_{x}x+\theta\right)  }\right]  $, $\left\langle \sigma
_{y}\right\rangle _{\left\vert x\right\vert <a}=s\left(  \left\vert
\mathcal{B}\right\vert ^{2}-\left\vert \mathcal{A}\right\vert ^{2}\right)
\cos\theta$, and $\left\langle \sigma_{z}\right\rangle _{\left\vert
x\right\vert <a}=\operatorname{Re}\left[  \mathcal{AB}^{\ast}e^{2ik_{x}%
x}\left(  e^{2i\theta}+1\right)  \right]  $, and in region III $\left\langle
\sigma_{x}\right\rangle _{x>a}=s\left\vert \mathcal{T}\right\vert ^{2}%
\sin\theta$, $\left\langle \sigma_{y}\right\rangle _{x>a}=-s\left\vert
\mathcal{T}\right\vert ^{2}\cos\theta$, and $\left\langle \sigma
_{z}\right\rangle _{x>a}=0$.

Furthermore, for the case of the steps grown along the $x$ ($\Gamma\mathtt{-}%
$K) direction, with the same method above to treat the wave functions
(\ref{wfy}), we could also get the coefficients $\overline{\mathcal{R}%
}\mathtt{=}\frac{\gamma}{M}$, $\overline{\mathcal{A}}\mathtt{=}\frac{\alpha
}{M}$, $\overline{\mathcal{B}}\mathtt{=}\frac{\beta}{M}$, and $\overline
{\mathcal{T}}\mathtt{=}\frac{\tau}{M}$, where $M$, $\alpha$, $\beta$, $\gamma
$, and $\tau$ have the same form shown above but with the parameters expressed
as
\begin{widetext}
\begin{equation}%
\begin{array}
[c]{l}%
a_{0}=\left[  -\left(  w+u\right)  \phi_{1,s}-si\phi_{2,s}e^{i\theta}\right]
e^{-ik_{y}a},\\
a_{1}=\left[  \left(  w-u\right)  \phi_{1,s}-si\phi_{2,s}e^{-i\theta}\right]
e^{ik_{y}a},\\
a_{2}=\left[  \left(  w-u\right)  \phi_{1,s}+si\phi_{2,s}e^{i\theta}\right]
e^{-ik_{y}a},\\
a_{3}=\left[  -\left(  w+u\right)  \phi_{1,s}+si\phi_{2,s}e^{-i\theta}\right]
e^{ik_{y}a},\\
c_{1}=\left[  \left(  w-2u\right)  \phi_{1,s}+si\phi_{2,s}e^{i\theta}\right]
e^{ik_{y}a},\\
c_{2}=\left[  -\left(  w+u\right)  \phi_{1,s}-si\phi_{2,s}e^{i\theta}\right]
e^{ik_{y}a},\\
c_{3}=\left[  \left(  w-2u\right)  \phi_{1,s}-si\phi_{2,s}e^{-i\theta}\right]
e^{-ik_{y}a},
\end{array}%
\begin{array}
[c]{l}%
b_{0}=\left[  -\phi_{1,s}+\left(  w-u\right)  si\phi_{2,s}e^{i\theta}\right]
e^{-ik_{y}a},\\
b_{1}=\left[  -\left(  w+u\right)  si\phi_{2,s}e^{-i\theta}-\phi_{1,s}\right]
e^{ik_{y}a},\\
b_{2}=\left[  -\left(  w+u\right)  si\phi_{2,s}e^{i\theta}+\phi_{1,s}\right]
e^{-ik_{y}a},\\
b_{3}=\left[  \left(  w-u\right)  si\phi_{2,s}e^{-i\theta}+\phi_{1,s}\right]
e^{ik_{y}a},\\
f_{1}=\left[  -\left(  w+2u\right)  si\phi_{2,s}e^{-i\theta}+\phi
_{1,s}\right]  e^{ik_{y}a},\\
f_{2}=\left[  \left(  w-2u\right)  si\phi_{2,s}e^{i\theta}-\phi_{1,s}\right]
e^{ik_{y}a},\\
f_{3}=\left[  -\left(  w+2u\right)  si\phi_{2,s}e^{-i\theta}-\phi
_{1,s}\right]  e^{-ik_{y}a}.
\end{array}
\tag{16}\label{14}%
\end{equation}
\end{widetext}
After obtaining these coefficients, one could get the LDOS confined in the
nanosteps by employing Eq. (\ref{DOS}) in the text.

\section*{ACKNOWLEDGEMENTS}

We acknowledge the support of Natural Science Foundation of China under Grants
No. 91321003, No. 11304009, No. 91230203, No. 11274049, and No. 11004013, the
National Basic Research Program of China (973 Program) under Grant No.
2009CB929103, and the General Financial Grant from the China Postdoctoral
Science Foundation under Grant No. 2012M520148.

\section*{AUTHOR CONTRIBUTIONS}

ZGF did the calculations. ZGF, PZ, MC, ZW, FWZ, and HQL analyzed the results.
ZGF and PZ wrote the paper. PZ and ZGF were responsible for project planning
and execution.

\section*{AUTHOR INFORMATION}

The authors declare no competing financial interests. Correspondence and
request or materials should be addressed to PZ (zhang\_ping@iapcm.ac.cn)

\end{document}